%%% FM 98-8; mp_arc #98-??? ; chao-dyn #9804044
%%%%%%%%%%          Plain TeX
%**start of header
\newcount\mgnf\newcount\tipi\newcount\tipoformule
\newcount\aux\newcount\piepagina\newcount\xdata
\mgnf=0
\aux=0           %1 produce aux
\tipoformule=1   %0 usa aux; 1 no (usa i simboli dati)
\piepagina=2     %0 =data e #par.#pag; 1=data e #pag; 2=#pag
\xdata=0         %0 data del giorno, 1 data fissa da \Di:
\def\Di{30 aprile 1998}

\ifnum\mgnf=1 \aux=0 \tipoformule =1 \piepagina=1 \xdata=1\fi
\newcount\bibl
%\bibl= ?               % 0= rif [XXX], 1= rif. numerici
\ifnum\mgnf=0\bibl=0\else\bibl=1\fi
\bibl=0

% Per poter cambiare a piacimento il formato dei riferimenti
% bibliografici in <nome>.tex:
%
% 1: citare nella forma esemplificata da \ref{B}{2}{20}}
%    ove XXX e' un simbolo per le iniziali e 2 distingue i lavori con
%    le stesse iniziali,  7 e' il numero SIMBOLICO del riferimento per XXX2.
%    Il numero 7 puo' essere ARBITRARIO e viene automaticamente
%    riaggiustato al momento della compilazione (vedi punto 4)
%
% 2: Se si sceglie \bibl=0 si cita nella forma [XXX2]; se si sceglie
%    \bibl=1 si cita nella forma [numero di ordine di prima citazione].
%
% 3: La bibliografia va scritta nella forma \def{\qqq}{<citazione>}
%    in ordine alfabetico per autore attribuendo un simbolo
%    qualsiasi al testo <citazione} (ad es \def\REFCIT{citazione} e
%    va seguita dal comando  \rif{XXX}{2}{\qqq}{0}
%    scritto all' inizio di una riga altrimenti bianca. L' ultimo {0}
%    e' un parametro inutile ed e' li solo provvisoriamente per memoria
%    (di tentativi in corso).
%
% 4: Per il riaggiustamento automatico (in ordine di citazione)
%    occorre compilare (ottenendo oltre ai soliti la scheda ausiliaria
%    ref.b) e poi si deve eseguire il
%    programma rif <nome> che (usando ref.b) produce fin.tex e <nome>.tex
%    con i riferimenti giusti
%    in \bibl=1 e la si ricompila e stampa. La scheda iniziale <nome>.tex
%    diventa <nome>.old.
\openout8=ref.b
\ifnum\bibl=0
\def\ref#1#2#3{[#1#2]\write8{#1@#2}}
\def\rif#1#2#3#4{\item{[#1#2]} #3}
\fi

\ifnum\bibl=1
\def\ref#1#2#3{[#3]\write8{#1@#2}}
\def\rif#1#2#3#4{}

\fi

\def\9#1{\ifnum\aux=1#1\else\relax\fi}
\ifnum\piepagina=0 \footline={\rlap{\hbox{\copy200}\
$\st[\number\pageno]$}\hss\tenrm \foglio\hss}\fi \ifnum\piepagina=1
\footline={\rlap{\hbox{\copy200}} \hss\tenrm \folio\hss}\fi
\ifnum\piepagina=2\footline{\hss\tenrm\folio\hss}\fi

\ifnum\mgnf=0 \magnification=\magstep0
\hsize=13.5truecm\vsize=22.5truecm \parindent=4.pt\fi
\ifnum\mgnf=1 \magnification=\magstep1
\hsize=16.0truecm\vsize=22.5truecm\baselineskip14pt\vglue5.0truecm
\overfullrule=0pt \parindent=4.pt\fi

\let\a=\alpha \let\g=\gamma \let\d=\delta
\let\e=\varepsilon  \let\h=\eta
\let\th=\vartheta\let\k=\kappa \let\l=\lambda \let\m=\mu \let\n=\nu
\let\x=\xi \let\p=\pi  \let\s=\sigma \let\t=\tau
 \let\f=\varphi\let\ch=\chi \let\ps=\psi \let\o=\omega
  \let\D=\Delta 
   \let\F=\Phi
  
{\count255=\time\divide\count255 by 60 \xdef\oramin{\number\count255}
\multiply\count255 by-60\advance\count255 by\time
\xdef\oramin{\oramin:\ifnum\count255<10 0\fi\the\count255}}
\def\ora{\oramin }

%\Di e' definito all' inizio
\ifnum\xdata=0
\def\data{\number\day/\ifcase\month\or gennaio \or
febbraio \or marzo \or aprile \or maggio \or giugno \or luglio \or
agosto \or settembre \or ottobre \or novembre \or dicembre
\fi/\number\year;\ \ora}
\else
\def\data{\Di}
\fi

\setbox200\hbox{$\scriptscriptstyle \data $}
\newcount\pgn \pgn=1
\def\foglio{\number\numsec:\number\pgn
\global\advance\pgn by 1} \def\foglioa{A\number\numsec:\number\pgn
\global\advance\pgn by 1}
\global\newcount\numsec\global\newcount\numfor \global\newcount\numfig
\gdef\profonditastruttura{\dp\strutbox}
\def\senondefinito#1{\expandafter\ifx\csname#1\endcsname\relax}
\def\SIA #1,#2,#3 {\senondefinito{#1#2} \expandafter\xdef\csname
#1#2\endcsname{#3} \else \write16{???? ma #1,#2 e' gia' stato definito
!!!!} \fi} \def\etichetta(#1){(\veroparagrafo.\veraformula) \SIA
e,#1,(\veroparagrafo.\veraformula) \global\advance\numfor by 1
\9{\write15{\string\FU (#1){\equ(#1)}}} \9{ \write16{ EQ \equ(#1) == #1
}}} \def \FU(#1)#2{\SIA fu,#1,#2 }
\def\etichettaa(#1){(A\veroparagrafo.\veraformula) \SIA
e,#1,(A\veroparagrafo.\veraformula) \global\advance\numfor by 1
\9{\write15{\string\FU (#1){\equ(#1)}}} \9{ \write16{ EQ \equ(#1) == #1
}}} \def\getichetta(#1){Fig.  \verafigura \SIA e,#1,{\verafigura}
\global\advance\numfig by 1 \9{\write15{\string\FU (#1){\equ(#1)}}} \9{
\write16{ Fig.  \equ(#1) ha simbolo #1 }}} \newdimen\gwidth \def\BOZZA{
\def\alato(##1){ {\vtop to \profonditastruttura{\baselineskip
\profonditastruttura\vss
\rlap{\kern-\hsize\kern-1.2truecm{$\scriptstyle##1$}}}}}
\def\galato(##1){ \gwidth=\hsize \divide\gwidth by 2 {\vtop to
\profonditastruttura{\baselineskip \profonditastruttura\vss
\rlap{\kern-\gwidth\kern-1.2truecm{$\scriptstyle##1$}}}}} }
\def\alato(#1){} \def\galato(#1){}
\def\veroparagrafo{\number\numsec}\def\veraformula{\number\numfor}
\def\verafigura{\number\numfig}
\def\geq(#1){\getichetta(#1)\galato(#1)}
\def\Eq(#1){\eqno{\etichetta(#1)\alato(#1)}}
\def\eq(#1){\etichetta(#1)\alato(#1)}
\def\Eqa(#1){\eqno{\etichettaa(#1)\alato(#1)}}
\def\eqa(#1){\etichettaa(#1)\alato(#1)}
\def\eqv(#1){\senondefinito{fu#1}$\clubsuit$#1\write16{No translation
for #1} \else\csname fu#1\endcsname\fi}
\def\equ(#1){\senondefinito{e#1}\eqv(#1)\else\csname e#1\endcsname\fi}
\openin13=#1.aux \ifeof13 \relax \else \input #1.aux \closein13\fi
\openin14=\jobname.aux \ifeof14 \relax \else \input \jobname.aux
\closein14 \fi \9{\openout15=\jobname.aux} \newskip\ttglue

%\font\dodicirm=cmr12\font\dodicibf=cmbx12\font\dodiciit=cmti12
%\font\titolo=cmbx12 scaled \magstep2
%\font\ottorm=cmr8\font\ottoi=cmmi8\font\ottosy=cmsy8
%\font\ottobf=cmbx8\font\ottott=cmtt8\font\ottosl=cmsl8\font\ottoit=cmti8
%\font\sixrm=cmr6\font\sixbf=cmbx6\font\sixi=cmmi6\font\sixsy=cmsy6

\font\titolo=cmbx10 scaled \magstep1
\font\ottorm=cmr8\font\ottoi=cmmi7\font\ottosy=cmsy7
\font\ottobf=cmbx7\font\ottott=cmtt8\font\ottosl=cmsl8\font\ottoit=cmti7
\font\sixrm=cmr6\font\sixbf=cmbx7\font\sixi=cmmi7\font\sixsy=cmsy7

\font\fiverm=cmr5\font\fivesy=cmsy5\font\fivei=cmmi5\font\fivebf=cmbx5
\def\ottopunti{\def\rm{\fam0\ottorm}\textfont0=\ottorm%
\scriptfont0=\sixrm\scriptscriptfont0=\fiverm\textfont1=\ottoi%
\scriptfont1=\sixi\scriptscriptfont1=\fivei\textfont2=\ottosy%
\scriptfont2=\sixsy\scriptscriptfont2=\fivesy\textfont3=\tenex%
\scriptfont3=\tenex\scriptscriptfont3=\tenex\textfont\itfam=\ottoit%
\def\it{\fam\itfam\ottoit}\textfont\slfam=\ottosl%
\def\sl{\fam\slfam\ottosl}\textfont\ttfam=\ottott%
\def\tt{\fam\ttfam\ottott}\textfont\bffam=\ottobf%
\scriptfont\bffam=\sixbf\scriptscriptfont\bffam=\fivebf%
\def\bf{\fam\bffam\ottobf}\tt\ttglue=.5em plus.25em minus.15em%
\setbox\strutbox=\hbox{\vrule height7pt depth2pt width0pt}%
\normalbaselineskip=9pt\let\sc=\sixrm\normalbaselines\rm}
\catcode`@=11
\def\footnote#1{\edef\@sf{\spacefactor\the\spacefactor}#1\@sf
\insert\footins\bgroup\ottopunti\interlinepenalty100\let\par=\endgraf
\leftskip=0pt \rightskip=0pt \splittopskip=10pt plus 1pt minus 1pt
\floatingpenalty=20000
\smallskip\item{#1}\bgroup\strut\aftergroup\@foot\let\next}
\skip\footins=12pt plus 2pt minus 4pt\dimen\footins=30pc\catcode`@=12
\let\nota=\ottopunti\newdimen\xshift \newdimen\xwidth \newdimen\yshift
\def\ins#1#2#3{\vbox to0pt{\kern-#2 \hbox{\kern#1
#3}\vss}\nointerlineskip} \def\eqfig#1#2#3#4#5{ \par\xwidth=#1
\xshift=\hsize \advance\xshift by-\xwidth \divide\xshift by 2
\yshift=#2 \divide\yshift by 2 \line{\hglue\xshift \vbox to #2{\vfil #3
\includegraphics{#4.ps} }\hfill\raise\yshift\hbox{#5}}} \def\8{\write13}

\def\didascalia#1{\vbox{\nota\0#1\hfill}\vskip0.3truecm}
\def\V#1{{\,\underline#1\,}}
\def\T#1{#1\kern-4pt\lower9pt\hbox{$\widetilde{}$}\kern4pt{}}
\let\dpr=\partial \let\io=\infty\let\ig=\int
\def\fra#1#2{{#1\over#2}}\let\0=\noindent
\def\guida{\leaders\hbox to 1em{\hss.\hss}\hfill}
\def\tende#1{\vtop{\ialign{##\crcr\rightarrowfill\crcr
\noalign{\kern-1pt\nointerlineskip} \hglue3.pt${\scriptstyle
#1}$\hglue3.pt\crcr}}} \def\otto{\
{\kern-1.truept\leftarrow\kern-5.truept\to\kern-1.truept}\ }

\def\pagina{\vfill\eject}\let\ciao=\bye

\def\st{\scriptscriptstyle}
\def\*{\vskip0.3truecm}

\def\eg{\hbox{\it e.g.\ }}

\def\ie{\hbox{\it i.e.\ }}

\def\fiat{{}}
\def\\{\hfill\break} \def\={{ \; \equiv \; }}
\def\Im{{\rm\,Im\,}}\def\Re{{\rm\,Re\,}} \def\sign{{\rm
sign\,}}\let\arctg=\atan

\def\annota#1{\footnote{${}^{\bf#1}$}}
\ifnum\aux=1\BOZZA\else\relax\fi
\ifnum\tipoformule=1\let\Eq=\eqno\def\eq{}\let\Eqa=\eqno\def\eqa{}
\def\equ{{}}\fi
\def\defi{\,{\buildrel def \over =}\,}

\def\1{\ifnum\mgnf=0\pagina\else\relax\fi}
\def\W#1{#1_{\kern-3pt\lower6.6truept\hbox to 1.1truemm
{$\widetilde{}$\hfill}}\kern2pt\,}
\def\Re{{\rm Re}\,}\def\Im{{\rm Im}\,}

\def\II{{\cal I}}

\def\TT{{\cal T}}

\def\aa{{\V \a}}\def\nn{{\V\n}}\def\AA{{\V A}}
\def\pps{{\V \ps}}
\def\oo{{\V \o}}\def\nn{{\V\n}}

\def\ndpr{{\kern1pt\raise 1pt\hbox{$\not$}\kern1pt\dpr\kern1pt}}
\def\Ndpr{{\kern1pt\raise 1pt\hbox{$\not$}\kern0.3pt\dpr\kern1pt}}

\def\OO{{\cal O}}
\fiat

\def\Val{{\rm Val}}\def\hdm{{\h^{-\fra12}}}\def\hdp{{\h^{\fra12}}}
%**end of header
\fiat
\centerline{\titolo Reminiscences on science at I.H.E.S.}
\centerline{\titolo A problem on homoclinic theory and a brief review}
\*\*

\centerline{\bf Giovanni Gallavotti}
\*
\centerline{\it Universit\`a di Roma ``La Sapienza'', Fisica}
\centerline{\Di}
\vskip1.truecm {\bf Abstract.} {\it On the occasion of the 40-th
anniversary of IHES I present a few scientific reminiscences: most of
my scientific life has been marked by my visits and I run through them
concluding with the analysis of a problem that originated during my
last visit. The problem is to develop a convergent perturbative
algorithm for the construction of the ``Eliasson's potential'' for the
stable and unstable manifolds of an invariant torus: and to study its
properties. A brief review follows.}

\*

\0{\bf\S1. Reminiscences.}
\numsec=1\numfor=1\*

In 1966 I came to IHES as a young ``professor'': I was in fact just a
Ph.D. student, by all standards. On the recommendation of Sergio
Doplicher I was taken in the group of Ph.D.'s of Daniel Kastler. Of
course I was fully aware of ``being out of place'': no time, however,
was spent on this fact and soon I was eagerly working with Salvador
Miracle--Sol\'e on problems posed by David Ruelle. He went away for a
few weeks and we met all possible difficulties so that we decided to
write down the details of a proof that the problem given to us was not
soluble: to realize, when the proof was complete, that we had in fact
solved it. This was extending to many body interactions the
Kirkwood--Salsburg equations (also known as ``cluster expansion'') for
lattice system: it was a tremendous encouragement. It was for me the
opening of a new world; that of perturbation theory: suddenly we had
in our hands a most powerful instrument. We began using it to prove,
with Dereck Robinson, that the Ising model had no phase transitions at
high temperature, uniformly in the density.\annota1{We realized later
that Dobrushin had preceded us.} At about the same time Jean Lascoux
arrived with several papers from the Russian school; we saw that
our work made them transparent and we could explain them around: in
particular the work of Minlos and Sinai on phase coexistence appeared
as a great achievement and as proof of the flexibility of perturbation
theory which could {\it even} yield the analysis of the deep two phase
region. I kept thinking to the matter until, quite a few years later,
I could understand the theory of the fluctuations of the interface in
the $2$--dimensional Ising model at low temperature.

Leaving IHES in 1968 (I mean after may) I had met so many people (for
instance Joel Lebowitz) and absorbed so many ideas and techniques that
I was mature enough to pose problems without the help of Ruelle or of
my collaborator and close friend Miracle--Sol\'e. Saying farewell to
Ruelle I saw on his blackboard (that he kept as black and powderless
as a blackboard can possibly be until, much later, he gave it up,
inexplicably) the Navier--Stokes equations: that he commented by
stating that he was abandoning statistical mechanics for fluid
mechanics ``that seemed more interesting''.

I brought along my copy of the fluids of Landau--Lifschitz and a few
days later, in the US, I started studying it as I was sure I would
soon need it. Even though I was discouraged by the difficulties (very
transparent in spite of the book style which presents every problem as
simple and as completely solved) I kept following Ruelle's work
getting updates on the occasions of my frequent visits to IHES. Until,
in 1973, I suddenly realized and was fascinated by his proposal of the
existence of a probability distribution associated with chaotic
motions in a conceptual generality comparable to the one, familiar in
equilibrium statistical mechanics, of the Boltzmann--Gibbs
distribution: this was, in my view, a development much more
significant than the (timely and necessary) critique of Landau's
Ptolemaic theory of fluid mechanics\annota{2}{Unfortunately Ptolemy
has become the villain of science: I do not share at all this
superficial conception and I consider his work as great as
possible. Hence a Ptolemaic theory is very respectable in itself and
one needs work to criticize it, if at all possible.}  stemming out of
the Ruelle--Takens paper.

Ruelle's viewpoint on strange attractors made them simple objects via
the Markov partitions and Sinai's theory of Anosov systems. But I
could not go further: nevertheless the problem (``how to obtain some
directly observable consequences'' from Ruelle's principle) remained
hunting me and I thought that it was the right approach and kept
lecturing on the subject every year since while attempting even some
numerical experiments\annota{3}{Learning programming on an
archeological computer located in a little room of IHES, which had a
marvelous object, a plotter, linked to it: if Oscar Lanford, then a
member of IHES, considered numerical investigation of chaotic motions
worth of devoting time to it {\it then} I should know too.} but mostly
following other people experiments.

In the meantime I tried to understand Quantum Field Theory (I had
cultivatied the feeling of a deep connection between renormalization
theory and the Kirkwood Salsburg equations since the work with
Miracle--Sol\'e): in one of my visits to IHES I had met Francesco
Guerra and in a few words, explaining his own basic work on Nelson's
approach, he managed to make suddenly clear what I considered until
then impenetrable (the works of Nelson and of Glimm--Jaffe and
renormalization theory). So I started thinking to the matter and, also
after a memorable lecture of Eckmann on the use of the cluster
expansion in field theory, I was able to see the connection between
the renormalization group of Wilson, constructive field theory and the
cluster expansion developing a new interpretation of the mathematical
works done until then in $2$ and $3$ dimensional field theory. I
exposed the results at IHES with several people around (there by
chance, because Gelfand and Sinai were visiting also) and I will never
forget Niko Kuiper calling me in the Director's office and telling me
that my talk had been successful: the first time I had been in that
office was when Leon Motchane, a few months after my arrival and a few
days after completing the first paper with Miracle--Sol\'e (before the
KS--equations), told me that I was given a 15\% raise in my salary.

After the work in field theory I kept visiting regularly IHES where I
was attracted also by the new young member Jurg Fr\"olich: the policy
of IHES started by Motchane of hiring young people at the highest rank
(``member''), in spite of the obvious risks, is in my view what made
the place so exceptional and so interesting and productive. But I was
never able to collaborate with Fr\"olich. Instead I worked with Henri
Epstein in spite of the age difference: together with Pierre Collet we
studied smooth conjugacy between flows on a surface of constant
negative curvature stimulated by opinions by other visitors that what
we wanted to do was ``proved'' to be impossible. The effervescent
atmosphere around Dennis Sullivan always frightened me: his synthetic
and powerful approach was in a way opposite to my nature. But of
course I was among those who were impressed and attracted by his view
on the theory of interval maps and Feigenbaum's theory and tried
(without success) to imitate it in the theory of the tori breakdown in
conservative systems: a subject into which I was drawn by Joel
Lebowitz' request, while we were both at IHES, to explain in a talk the
KAM theory that I had boasted, in refereeing a paper for him, to have
understood from Arnold's celestial mechanics paper (the seminar took
place a few days later at the Ecole Politechnique).

A long gap in my visits followed: which more or less coincided with a
period in which I did not really work on new problems but looked at
consequences of previous ones. I kept nevertheless thinking to the old
questions and in my next long visit, in 1993, I was ready to attack a
problem that E.G.D. Cohen (also a visitor at the time) proposed me to
jointly work to find the connection between the experiments and
theoretical ideas that he and his coworkers had been developing, in
parallel with the works of Hoover and coworkers, on nonequilibrium
statistical mechanics. Two exciting years followed: in retrospect all
that could have been done in one afternoon; but for us it was very
hard work and at the end of the two years we could make a precise
proposal (the ``chaotic hypothesis'') for the application of Ruelle's
principle to some simple but concrete dynamical problems. My last
visit to IHES was in 1997: I went there with bellicose projects to
continue working on nonequilibrium statistical mechanics. But as usual
I was taken away by other intervening projects\annota{4}{One was not a project
but nevertheless took a lot of time: understanding the basics of
Linux.} which ended having to do with the theory of homoclinic orbits
and the transversality of their intersections: for this reason,
although this is not a problem of the same size of the previously
mentioned ones, I will devote to it the technical part of this note. I
am sure that, as all problems that I started at IHES, this too will be
fruitful and linked tightly with the previous works on the cluster
expansion.

I also must remember here M.me Annie Rolland-Motchane: she, the first
general secretary to IHES, clearly shared the merit of conceiving the
structure of IHES and of making it work.

It is a structure that during the last 33 years always made me feel
that IHES was an ideal place as a source of inspiration and of
systematic work and it has been for me a privilege to be able to work
there. I took this literally, as privileges must be earned: I was so
absorbed in my work, taking up most of the nights (the day being
reserved to wandering around trying to get the most out of the people
present, with a few trips to the small library just to remind myself
that even though it was small it was still completely full of things
that I ignored): to the point that I could always manage avoiding the
frustration due to my feeling of being there out of place among colleagues
far more knowledgeable than I could possibly hope to be.
\*

\0{\bf\S2. Homoclinic intersections in Hamiltonian systems. A ``field
theoretic'' approach.}
\numsec=2\numfor=1\*

Consider a Hamiltonian system:

$$H=\oo\cdot\AA+\fra{I^2}{2J_0} + J_0 g^2 \,(\cos \f-1) +\e
f(\aa,\f)\Eq(2.1)$$
where $\oo=(\o_1,\o_2)\in R^2,\,\AA=(A_1,A_2)\in R^2$,
$\aa=(\a_1,\a_2)\in T^2$, $I\in R$, $\f\in T^1$ and $f(\aa,\f)$ is an
even analytic function of the angles $\aa,\f$ which is a trigonometric
polynomial in $\f$; its analyticity domain is supposed to be $|\Im
\a_j|<\x$. Therefore \equ(2.1) represents a quasi--periodically forced
pendulum.

What follows can be easily extended to the system obtained from
\equ(2.1) by adding to it $\fra12(J^{-1}\AA,\AA)$ where $J$ is a positive
diagonal matrix. The case \equ(2.1) is the limiting case as $J\to+\io$
of the latter extension. The more general model is called {\it
anisochronous} or {\it Thirring's model} as its peculiar properties
were pointed out by Thirring, [T].

Supposing that $\oo$ verifies a Diophantine property
$|\oo\cdot\nn|>C^{-1}|\nn|^{-1}$ for all $\nn\in Z^2$, integer
components {\it nonzero} vectors, it follows that for $\e$ small
enough and for each $\AA_0$ there is an invariant torus $\TT(\AA_0)$
which has equations parameterized by $\pps\in T^2$ of the form
$\AA=\AA_0+\V H(\pps)$, $\aa=\pps+\V h(\pps)$,
$I=R(\pps),\f=S(\pps)$. The motion on $\TT(\AA_0)$ is $\pps\to\pps+\oo
t$ and the average of $\V H$ over $\pps$ (hence on time) vanishes.

Hence we have a natural parameterization of the invariant tori by the
(time) average value $\AA_0$ of their action variables. In the more
general Thirring's model there is also a relation between the rotation
vector $\oo$ on an invariant torus and the (time) average action
$\AA_0$ of the motion on it: namely $\oo=J^{-1}\AA_0$ (this is a
remarkable property which prompted the appellative of ``twistless''
for such invariant tori).

Let $E(\AA_0)$ be the energy of the motions on the invariant torus
$\TT(\AA_0)$.  The torus $\TT(\AA_0)$ is unstable and its stable and
unstable manifolds $W^u(\AA_0),W^s(\AA_0)$ will consist of points with
the same energy $E(\AA_0)$; the intersection of $W^a(\AA_0)$, $a=u,s$,
with the plane $\f=\p$ and with the energy surface $H=E(\AA_0)$
consists of points that can denoted as:
$X(\aa)=(\p,\aa,I^a(\aa),\AA^a(\aa))$, for $a=u,s$ and $\aa\in
T^2$. In general we denote a point in phase space $X=(\f,\aa,I,\AA)$.

Therefore on the $4$--dimensional ``section'' $\f=\p, H=E(\AA_0)$ the
manifolds have dimension $2$. The vector $\V
Q(\aa)=\AA^s(\aa)-\AA^u(\aa)$ defines the {\it splitting}. It has been
remarked by Eliasson that the vector $\V Q(\aa)$ is a {\it gradient}:
this means that there is a function $\F(\aa)$ such that $\V
Q(\aa)=\V\dpr_\aa \F(\aa)$. I will call $\F(\aa)$ the {\it Eliasson's
potential}.

Hence, since $\aa$ varies on $T^2$, there will be at least one point
$\aa_0$ where $\V Q(\aa_0)=\V0$. The corresponding point $X(\aa_0)$
will be, of course, {\it homoclinic}, \ie common to both manifolds.
In the case at hand the parity properties of $f$ imply that
$\aa_0=\V0$ is one such point, see for instance [G3].

It follows from the general theory of the tori $\TT(\AA_0)$, due to
Graff, which is very close to KAM theory (particularly easy in the
case \equ(2.1), see for instance [G2]; see \S5 of [CG] for the general
anisochronous case), that the function $\F(\aa)$ is analytic in $\e$
for $\e$ small, and such are the functions $\AA^a(\aa), I^a(\aa)$, for
$a=u,s$.

Therefore we must be able to find the coefficients of their power
series expansion in $\e$: in fact the expansion of $\AA^a(\aa),
I^a(\aa)$ has been thoroughly discussed in all details in [G3], in the
case of a polynomial $f$. It would be easy to derive from it the
expansion for $\F(\aa)$.

However here, \S3, I will give a {\it self contained} exposition which
leads to the expansion for $\F(\aa)$ borrowing from [G3] only a few
algebraic identities that it would be pointless to prove again. And I
will dedicate \S4 to a brief review of the results that are known (to
me) on the matter or that can be easily derived from existing papers
(which usually do not deal with Eliasson's potential but rather with
its gradient). A question will be formulated at the end of this
work.

The expansion below is the best tool, to my knowledge, to achieve a
unified proof of various theorems dealing with the {\it homoclinic
splitting}: the latter will be defined here as the Hessian determinant
$\D$ of $\F(\aa)$ evaluated at the homoclinic point at $\aa=\V0$ (in
our case).  Evaluation of the latter determinant was begun by Melnikov
who gave a complete solution (in spite of claims in other directions)
in the ``simple'' case in which $\oo$ is regarded as fixed. The
quasi--periodic case was treated in the spirit of the present work,
\ie with attention to the Arnold's diffusion problem, in [HM]
(togheter with the important realization of the usefulness of improper
integrations analysis).

Considerable interest has been dedicated to the problem by various
authors. The results are not easily comparable as every author seems
to give his own definition of splitting: the most remarkable results
have been developed in the russian school approach (based on the key
works of Melnikov, Neishtadt, Lazutkin, [Ge]). And sometimes it has
even been difficult to realize that some papers were just plainly
incorrect (like the result in \S10 of [CG]).

The interest of the above definition is its direct relation with the
problem of showing the existence of {\it heteroclinic strings}, \ie
sequences of tori $\TT(\AA_i)$ such that $W^u(\AA_i)\cap
W^s(\AA_{i+1})\ne\emptyset$. Given a curve $\ell\to\AA(\ell)$, with
$\TT(\AA(\ell))$ being equienergetic tori (\ie such that
$\oo\cdot\AA(\ell)=const$ in the simple case \equ(2.1), as one
realizes from symmetry considerations), and calling $\AA^a(\aa,\ell),
\, a=u,s$, the equations of the manifolds of $\TT(\AA(\ell))$, one
finds such a string if one can show that $\V
A^u(\aa;\ell)-\AA^s(\aa;\ell')=\V0$ admits a solution for $\ell'$
close enough to $\ell$. Therefore one has to apply the implicit
functions theorem around $\aa=\V0$ and $\ell'=\ell$: the determinant
of the Jacobian matrix controlling this problem is, clearly, the
splitting defined above. Hence proving that the splitting is nonzero
implies the existence of heteroclinic strings.

In the anisochronous case not all the $\AA(\ell)$ are necessarily the
average actions of an invariant torus (because of the resonances
always present when the rotation vectors do depend on the actions) so
that it becomes important to measure the size of the splitting
compared to the size of the ``gaps'' on the curve parameterized by $\ell$.
Hence we need to know quite well the dependence on $\oo$ of the
splitting and of the gaps and their relative sizes, see [GGM2].

I know of no paper in which the above definition of splitting is used
in the case of quasi periodic forcing models, other than [CG] and the
later [G3], [GGM1], [GGM2], [GGM3] or the related [BCG]: I will not
discuss the papers using other definitions.

\*
\0{\bf\S3.  Feynman's graphs for Eliasson's potential.}
\numsec=3\numfor=1\*

We set $J_0=1$ for simplicity.  Let $a=u,s$ and
$X^a(0)=(\p,\aa,I^a(\aa),\AA^a(\aa))$; and let
$X^a(t)=(\f^a(\aa,t),\aa+\oo t, I^a(\aa,t),\AA^a(\aa,t))$ be the
solution of the equations of motion for \equ(2.1): $\dot\f=I,\,
\dot\aa=\oo,\, \dot I=-\dpr_\f f_0(\f)\,-\e\dpr_\f f_1(\aa,\f),\,
\dot\AA=-\e\dpr_\aa f_1(\aa,\f)$ where $f_0=g^2\cos\f$ and
$f_1=f(\aa,\f)$. In the case $\e=0$ the stable and unstable manifolds
of the torus with (average) action $\AA_0$ coincide and the parametric
equations of their stable and unstable manifolds are at $\f=\p$ simply
$X=(\f=\p, \aa,I=-2g,\AA=\AA_0)$ and the motion are
$X^{(0)}(t)=(\f^0(t)=4\arctg e^{-gt},\aa+\oo
t,I(t)=-g\sqrt{2(1-\cos\f^0(t))}, \AA=\AA_0)$.

Hence if $X^a(\aa,t)=X^{(0)}(\aa,t)+\e X^{a,(1)}(\aa,t)+\e^2
X^{a,(2)}(\aa,t)+\ldots$ we can immediately write the equations for
$X^{(k)}= (\f^{a,(k)},\V0,I^{a,(k)}\AA^{a,(k)})$, $k\ge1$, where the
angle components vanish identically because the equations for $\aa$
are trivially solved by the order $0$ solution. The latter is a
property of the isochrony of \equ(2.1) and it does not hold in the
anisochronous case, which is however equally easy to treat, see [G3].
The equations are, dropping for simplicity the label $a$:

$$\eqalignno{
\pmatrix{\dot \f^{(k)}\cr\dot I^{(k)}\cr}=&\pmatrix{0&1\cr
g^2\cos\f^0&0\cr}\cdot\pmatrix{\f^{(k)}\cr I^{(k)}\cr}+\pmatrix{0\cr
F_0^{(k)}\cr}=
L(t)\cdot\pmatrix{\f^{(k)}\cr I^{(k)}\cr}+\pmatrix{0\cr
F_0^{(k)}\cr}\cr
\dot\AA^{(k)}=&\V F^{(k)}&\eq(3.1)\cr}$$
and the functions $F_0,\V F$ are deduced from the equations of motion;
recalling that $f_0=g^2 \cos\f$ and $f_1=f(\aa,\f)$:

$$\eqalign{
F_0^{(k)}=&\sum_{h=2}^k -\fra{1}{h!}\dpr^{h+1}_\f
f_0(\f^0(t))\sum_{k_1+\ldots+k_h=k\atop k_j\ge1}\prod_{j=1}^h
\f^{(k_j)}(t)+\cr
&+\sum_{h=1}^k-\fra{1}{h!}\dpr_\f^{h+1} f_1(\aa+\oo
t,\f^0(t))\sum_{k_1+\ldots+k_h=k-1\atop k_j\ge1}
\prod_{j=1}^h\f^{(k_j)}(t)\cr
\V F^{(k)}=&\sum_{h=1}^k-\fra{1}{h!}\dpr_\aa \dpr^h_\f f_1(\aa+\oo
t,\f^0(t))\sum_{k_1+\ldots+k_h=k-1\atop k_j\ge1}
\prod_{j=1}^h\f^{(k_j)}(t)\cr
}\Eq(3.2)$$
It is convenient to rewrite the above expressions in a more synthetic
and symmetric form:

$$\eqalign{
F_0^{(k)}=&\sum_{\d=0,1}\sum_{h=2-\d}^k -\fra{1}{h!}\dpr^{h+1}_\f
f_\d(\aa+\oo t,\f^0(t))\sum_{k_1+\ldots+k_h=k-\d\atop
k_j\ge1}\prod_{j=1}^h
\f^{(k_j)}(t)\cr
\V F^{(k)}=&\sum_{\d=0,1}\sum_{h=2-\d}^k-\fra{1}{h!}\dpr_\aa
\dpr^h_\f f_\d(\aa+\oo
t,\f^0(t))\sum_{k_1+\ldots+k_h=k-\d\atop k_j\ge1}
\prod_{j=1}^h\f^{(k_j)}(t)\cr}\Eq(3.3)$$
where of course in the second relation only the $\d=1$ term does not
vanish.

Hence if $W(t)=\pmatrix{w_{00}(t)&w_{01}(t)\cr w_{10}(t)&w_{11}(t)\cr}$
is the solution of the $2\times2$--matrix equation $\dot W=L(t)W,
W(0)=1$ we find: $\pmatrix{\f^{(k)}\cr I^{(k)}\cr}=
W(t)\Big(X^{(k)}(0)+\ig_0^t W(\t)^{-1}\pmatrix{0\cr F_0^{(k)}(\t)\cr}
d\t\Big)$.

The matrix $W(t)$ can be easily explicitly computed: the matrix
elements are holomorphic for $g|\Im t|<\fra\p2$ and $w_{00},w_{10}$
tend to zero as $t\to\pm\io$ as $e^{-g|t|}$ while the other column
elements tend to $\io$ as $e^{g|t|}$. We shall only need the matrix
elements $w_{00},w_{01}$ which have a simple pole at $\pm
i\fra{\p}{2g}$. It is $w_{00}(t)=\fra1{\cosh gt}$ and
$w_{01}(t)=\fra1{2g}(\fra{gt}{\cosh gt}+\sinh gt)$.

Since we need only $\f^{(h)}$, $h=1,\ldots,k-1$, to evaluate $F^{(k)}$
we spell out only the expressions of $\f^{(k)}(t)$ and $\AA^{(k)}(t)$:

$$\eqalign{
\f^{(k)}(t)=&w_{01}(t)\,\Big(I^{(k)}(0)+\ig_0^t w_{00}(\t)
F^{(k)}(\t) d\t\Big) -w_{00}(t)\ig_0^t w_{01}(\t) F^{(k)}(\t) d\t\cr
\AA^{(k)}(t)=&\AA^{(k)}(0)+\ig_0^t \V F^{(k)}(\t) d\t\cr}
\Eq(3.4)$$
We want to determine, for each choice of $\aa$, the initial data
$I^{(k)}(0),\AA^{(k)}(0)$ so that the motion is asymptotically quasi
periodic, because we want to impose that the motion {\it tends} to the
invariant torus $\TT(\AA_0)$. This means that the initial data must be
determined in a different way depending on whether we impose this
condition as $t\to+\io$ or as $t\to-\io$: in the first instance we
determine the stable manifold and in the second the unstable one.

This condition is imposed simply by requiring
$\AA^{(k)}(0)=-\ig_0^{\s\io} \V F^{(k)}(\t)d\t$ where $\s=+$ if we
impose the condition at $t=+\io$ or $\s=-$ if we impose the condition
at $t=-\io$ and, likewise, $I^{(k)}(0)=-\ig_0^{\s\io} w_{00}(\t)
F_0^{(k)}(\t)d\t$ {\it provided the integrals converge}. Therefore
\equ(3.4) become:

$$\eqalign{
\f^{(k)}(t)=&\ig_{\s\io}^t\Big(w_{01}(t)w_{00}(\t)-w_{00}(t)w_{01}(\t)\Big)
F_0^{(k)}(\t)\,d\t\,- \cr&-
w_{00}(t)\ig_{\s\io}^0 w_{01}(\t)F_0^{(k)}(\t)\,d\t \cr
\AA^{(k)}(t)=&\ig_{\s\io}^t \V F^{(k)}(\t)\,d\t\cr}\Eq(3.5)$$
and we have a simple recursion relation {\it provided the integrals converge}.

The above expressions however may involve non convergent integrals and
eventually they really do, in general. This has the consequence that (for
instance) it is by no means true that
$\AA^{s,(k)}(t)\tende{t\to+\io}0$ as the \equ(3.5) would imply if the
integrals were proper. Of course there will be no ambiguity about the
meaning of such {\it improper} integrals. The meaning is uniquely
determined simply by the requirement that the asymptotic form of
\equ(3.4) be a quasi periodic function.

Since the functions in the integrands can always be written as series
of functions of the form $\s^\ch\,\fra{(\s t g)^j}{j!} e^{-g\s \t h}
e^{i\oo\cdot\nn \, t}$ for some $\ch=0,1,j,h,\nn$ with $\s=\sign \t$,
it is easy to see that the rule for the evaluation of the integrals of
such function is simply that of introducing a cut off factor $e^{-R\s
\t}$ with $\Re R$ large enough so that all the integrals converge,
then one performs the (now convergent) integrals and then one takes
the residue at $R=0$ of the result: see \S3 of [G3] for a more
detailed discussion. To derive this rule one shoould try a few simple
cases (like evaluating the second order explitcitly).

Improper integrals are very familiar in perturbation theory of quantum
fields where they are normally introduced to obtain compact and
systematic representations of the coefficients of perturbation
expansions. Typically a Feynman diagram value is given by an improper
integral: the algorithm is so familiar that it has become usual not to
even mention which are the rules for the evaluation of such integrals.

Since the rules for the evaluation of the above imoroper integrals
are discussed in detail in [G3] I shall not dwell on them and,
instead, I proceed immediately to use the improper integrals in the
same way they are used in quantum field theory: \ie to find a simple
diagrammatic representation of the iterative scheme described
above. It is remarkable that such a scheme was found by Eliasson in
his breakthrough theory of the KAM series, [E], without any reference
to field theory: he has independently developed a diagrammatic
representation of the KAM series.

We represent $\f^{(k)}(t)$ as:

\eqfig{25pt}{0pt}{\ins{15pt}{8pt}{$t$}\ins{33pt}{3pt}{$(k)$}}{kfig1}{Fig.1}

and, with the same ``logic'':

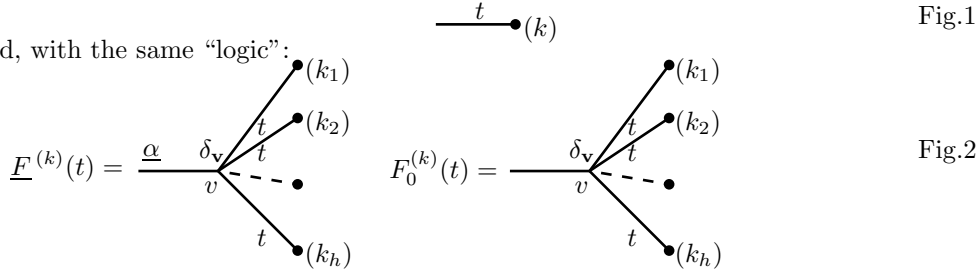
\eqfig{250pt}{70pt}{
%\ins{155pt}{40pt}{$t$}
\ins{163pt}{42pt}{$\bf\d_v$}
\ins{185pt}{50pt}{$t$}
\ins{185pt}{40pt}{$t$}
\ins{185pt}{8pt}{$t$}
\ins{203pt}{73pt}{$(k_1)$}
\ins{203pt}{53pt}{$(k_2)$}
\ins{203pt}{3pt}{$(k_h)$}
\ins{95pt}{38pt}{$F_0^{(k)}(t)=$}
\ins{165pt}{26pt}{$v$}
\ins{-50pt}{38pt}{$\V F^{(k)}(t)=$}
\ins{0pt}{40pt}{$\bf\aa$}
\ins{25pt}{26pt}{$v$}
%\ins{15pt}{40pt}{$t$}
\ins{23pt}{42pt}{$\bf\d_v$}
\ins{45pt}{50pt}{$t$}
\ins{45pt}{40pt}{$t$}
\ins{45pt}{8pt}{$t$}
\ins{63pt}{73pt}{$(k_1)$}
\ins{63pt}{53pt}{$(k_2)$}
\ins{63pt}{3pt}{$(k_h)$}
}{kfig2}{Fig.2}
\*

\0where it is $\sum_j k_j=k-\d_v$, see \equ(3.3); the label $\d_v$ can
be $0$ or $1$: the first drawing represents the term with $\d=\d_v$ in
the expression for $\V F^{(k)}$ in \equ(3.3), and the second drawing
represents the contribution to $F_0^{(k)}$ with $\d=\d_v$.

The node $v$ represents $-\dpr^{h+1}_\f f_{\d_v}$ times $\fra1{h!}$ in
the second graph and $-\dpr_\aa\dpr^h_\f f_1$ times $\fra1{h!}$ in the
first. Because of the $\dpr_\aa$ derivative we can imagine that in the
first graph thelabel $\d_v$ on the node $v$ is {\it constrained} to be
$1$.

We can in the same way represent $\f^{(k)}(t)$ and $\AA^{(k)}(t)$: we
can in fact change the labels $t$ on the lines merging into the node
$v$ into labels $\t$ and interpret the node $v$ as representing an
integration operation over the time $\t$; one get in this way
the following graphs:
\*
\eqfig{250pt}{70pt}{
\ins{155pt}{40pt}{$t$}
\ins{163pt}{42pt}{$\bf\d_v$}
\ins{185pt}{50pt}{$\t$}
\ins{185pt}{40pt}{$\t$}
\ins{185pt}{8pt}{$\t$}
\ins{203pt}{73pt}{$(k_1)$}
\ins{203pt}{53pt}{$(k_2)$}
\ins{203pt}{3pt}{$(k_h)$}
\ins{95pt}{38pt}{$\f^{(k)}(t)=$}
\ins{165pt}{26pt}{$v$}
\ins{0pt}{40pt}{$\aa$}
\ins{-50pt}{38pt}{$\AA^{(k)}(t)=$}
\ins{25pt}{26pt}{$v$}
\ins{15pt}{40pt}{$t$}
\ins{23pt}{42pt}{$\bf \d_v$}
\ins{45pt}{50pt}{$\t$}
\ins{45pt}{40pt}{$\t$}
\ins{45pt}{8pt}{$\t$}
\ins{63pt}{73pt}{$(k_1)$}
\ins{63pt}{53pt}{$(k_2)$}
\ins{63pt}{3pt}{$(k_h)$}
}{kfig2}{Fig.3}
\*
The node $v$ with the label $\d_v$, which we noted that it must be $1$
in the first drawing and that can be either $0$ or $1$ in the second,
has to be thought as representing the operations acting on a generic function
$F$:

$$\eqalignno{\II_\s F(t)=&\ig_{\s\io}^t F(\t) \, d\t&\eq(3.6)\cr
\OO_\s F(t)=&\ig_{\s\io}^t\Big(w_{01}(t)w_{00}(\t)-w_{00}(t)w_{01}(\t)\Big)
F(\t)\,d\t\,- \,
w_{00}(t)\ig^{\s\io}_0 w_{01}(\t)F(\t)\,d\t \cr}$$
where $\s=+$ if we study the stable manifold and $\s=-$ if we study
the unstable one.

In this way the graphs of Fig.3 represent respectively the {\it values}:

$$
\fra1{h!}\II_\s\Big(-\dpr_\aa\dpr^h_\f f_1[\t] \prod_{j=1}^h
\f^{(k_j)}[\t]\Big)(t),\qquad
\fra1{h!}\OO_\s\Big(-\dpr^{h+1}_\f f_\d[\t]\prod_{j=1}^h
\f^{(k_j)}[\t]\Big)(t))x\Eq(3.7)$$
where an argument in square brackets means a dummy integration
variable, inserted just as a reminder of the integration operation
involved; furthermore $f_\d[t]$ abbreviates $f_\d(\aa+\oo t,\f^0(t))$.

Clearly $I^{(k)},\AA^{(k)}$ can be expressed simply by summing over
the labels $k_j$ and $\d$ the values of the graphs in Fig.3: the
summations should run over the same ranges appearing in \equ(3.2), \ie
$h$ between $2-\d$ and $k$, and $k_j\ge1$ such that $\sum_j k_j=k-\d$
and $\d=0,1$). If we study the stable manifold we must take $\s=+$ and
if we study the unstable one me must take $\s=-$ and
$I^{(k)},\AA^{(k)}$ become respectively $I^{s,(k)},\AA^{s,(k)}$ or
$I^{u,(k)},\AA^{u,(k)}$.

It is now immediate to iterate the above representation; one simply
recalls that each symbol:

\eqfig{25pt}{0pt}{\ins{15pt}{8pt}{$t$}\ins{33pt}{3pt}{$(k_j)$}}{kfig1}{Fig.4}
\*
\0represents $\f^{(k_j)}(t)$ and that \equ(3.7) is multi linear in the
$\f^{(k_j)}(t)$. This leads to representing $\AA^{(k)}(t)$ as sum of
values of graphs $\th$ of the form:

\*

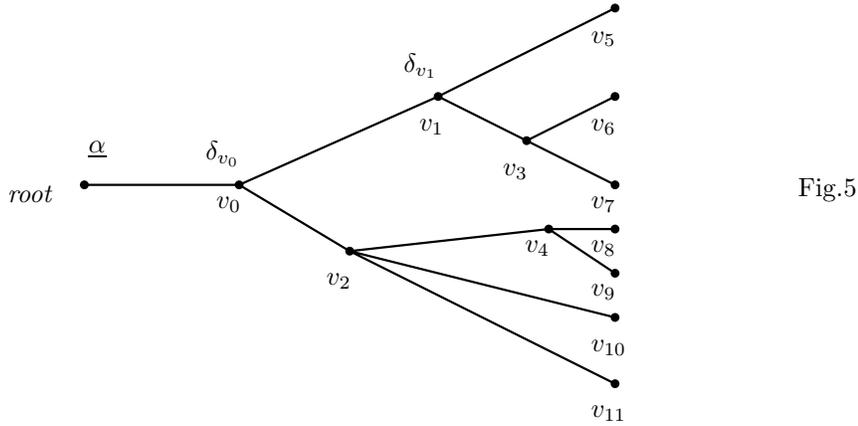
\eqfig{199.99919pt}{141.666092pt}{
\ins{-29.16655pt}{74.999695pt}{\it root}
\ins{0.00000pt}{91.666298pt}{$\bf\aa$}
\ins{49.99979pt}{70.833046pt}{$v_0$}
\ins{45.83314pt}{91.666298pt}{$\d_{v_0}$}
\ins{126.66615pt}{99.999596pt}{$v_1$}
\ins{120.83284pt}{124.999496pt}{$\d_{v_1}$}
\ins{91.66629pt}{41.666500pt}{$v_2$}
\ins{158.33270pt}{83.333000pt}{$v_3$}
\ins{191.66589pt}{133.332794pt}{$v_5$}
\ins{191.66589pt}{99.999596pt}{$v_6$}
\ins{191.66589pt}{70.833046pt}{$v_7$}
\ins{191.66589pt}{-8.333300pt}{$v_{11}$}
\ins{191.66589pt}{16.666599pt}{$v_{10}$}
\ins{166.66600pt}{54.166447pt}{$v_4$}
\ins{191.66589pt}{54.166447pt}{$v_8$}
\ins{191.66589pt}{37.499847pt}{$v_9$}
}{bggmfig0}{\hskip.6truecm Fig.5}
\kern0.9cm
\didascalia{A graph $\th$ with $p_{v_0}=2,p_{v_1}=2,p_{v_2}=3,
p_{v_3}=2,p_{v_4}=2$ and $k=12$, and some labels. The lines length is
drawn of arbitrary size. The nodes labels $\d_v$ are indicated only
for two nodes. The lines are imagined oriented towards the root and
each line $\l$ carries also a (not marked) label $\t_v$, if $v$ is the
node to which the line leads; the root line carries the label $t$ but
its ``free'' extreme, that we call the ``root'', is not regarded as a
node.}

The meaning of the graph is recursive: all nodes $v$, see Fig.5,
represent $\OO_\s$ operations except the ``first'' node $v_0$ which
instead represents a $\II$ operation; the extreme of integration is
$+\io$ if we study the stable manifold and $-\io$ if we stufy the
unstable one. Furthermore each node represents a factor
$-\fra1{p_v!}\dpr^{p_v+1}_\f f_{\d_v}[\t_v]$ if $p_v$ is the number of
lines merging into $v$, {\it except} the ``first'' node $v_0$ which
represents $-\dpr_\aa\dpr^{p_v}_\f f_{\d_v}[\t_v]$ instead. The
product $\prod_v \fra1{p_v!}$ is the ``combinatorial factor'' for the
node $v$

The lines merging into a node are regarded as distinct: \ie we imagine
that they are labeled from $1$ to $p_v$, but we identify two graphs
that can be overlapped by permuting suitably and independently
the lines merging into the nodes.

It is more convenient to think that all the lines are numbered from
$1$ to $m$, if the graph has $m$ lines, still identifying graphs that
can be overlapped under the above permutation operation (including the
line numbers). In this way a graph with $m$ lines will have a
combinatorial factor simply equal to $\fra1{m!}$ provided we define
$1$ instead of $\prod_v \fra1{p_v!}$ the combinatorial factor of each
node: {\it we shall take the latter numbering option}. Hence in Fig.5
one has to think that each line carries also a number label although
the line numbers, distinguishing the lines, are not shown.

The endnodes $v_i$ should carry a $(k_i)$ label: but clearly unless
$k_i=1$ they would represent a $\f^{(k_i)}$ which could be further
expanded; hence the graphs in Fig.5 should have the labels $(k_i)$
with $k_i=1$: this however carries no information and the labels are
not drawn. The interpretation of the endnodes is easily seen that has
to be: $\OO_\s(-\dpr_\f f_{\d_{v_i}})(\t_{v'_i})$ if $v'_i v_i$ is the
line linking the endnode $v_i$ to the rest of the graph. An exception
is the trivial case of the graph with only one line and one node: this
represents $\II(-\dpr_\aa f_{\d_{v_0}})(t)$ and it will be called the
{\it Melnikov's graph}.

In this way we have a natural decomposition of $\AA^{(a,(k)}(\aa,t)$
as a sum of {\it values} of graphs. It is now easy to represent the
power series expansion of the trajectories on the manifolds
$W^a(\TT(\AA_0))$: one simply collects all graphs with labels $\d_v$
with $\sum_vd_v=k\ge1$ (they can have at most $2k$ lines, if one
looks at the restrictions on the labels) and adds up their ``values''
obtaining the coefficient $\AA^{a,(k)}(t)$. The $\OO_\s$ and $\II_\s$
operations involve integrals with $\s\io$ as an extreme and one has,
obviously, to choose $\s=+$ if $a=s$ and $\s=-$ if $a=u$.

Since all the integration operations $\OO$ or $\II$ are, in general,
improper we see the convenience of the graphical representation and
its analogy with the Feynman graphs of quantum field theory: in fact
this is {\it more than an analogy} as the above graphs can be regarded
as the Feynman graphs of a suitable field theory: see [GGM0] for the
discussion of a similar case (\ie the KAM theory representation as a
quantum field theory).

An essential feature is missing: namely the graphs have no loops (they
are in fact tree graphs). This major simplification is compensated by
the major difficulty that the number of lines per node is {\it
unbounded} (\ie a field theory that generated the graphs would have to
be ``non polynomial'').

Noting that the value of each graph is a function of $\aa$ we now have
to check that each $\AA^{a,(k)}(0)$ has the form
$\AA^{a,(k)}(0)=\dpr_\aa \F^{a,(k)}$.

For this purpose we consider graphs like Fig.5 but with the root
branch {\it deleted} keeping however a {\it mark} on the first node $v_0$ to
remember that the line has been taken away. We call such a graph a
{\it rootless} graph.

{\it It is convenient to define the value $\Val_\s(\th)$ of such
rootless trees}: its is defined as before but the marked node now
represents the operation $\II_{\s,0}(F)\=\ig_{\s\io}^0 d\t F(\t)$ with
$\s=+$ for the analysis of the stable manifold and $\s=-$ for the
unstable, and the function $\dpr^{p_{v_0}} f_{\d_{v_0}}[\t]$ (keeping
in mind that the marked node {\it must} have $\d_{v_0}=1$, by
construction).

The key remark is now the identity (``Chierchia's root
identity'', see [G3]):

$$\II_{\s,0}( F\, \OO_\s(G))=\II_{\s,0}( G \,\OO_\s(F))\Eq(3.8)$$
which is an algebraic identity as our improper integrals only involve
functions $F, G$ linear combinations of ``monomials'' of the form
$\s^\ch\,\fra{(\s t g)^j}{j!} e^{-g\s \t h} e^{i\oo\cdot\nn \, t}$ for
some $\ch=0,1,j,h,\nn$ with $\s=\sign \t$, see above, for which both
sides of \equ(3.7) can be explicitly and easily evaluated.

This identity can be used to relate the values of different graphs: it
means that the values of two rootless tress differing only because the
mark is on different nodes and otherwise superposable are {\it
identical}: this can be seen easily by successive applications of the
identity \equ(3.7), see [G3].

Therefore if we define:

$$\F^{\s,(k)}(\aa)=\fra1k \sum_\th \Val_\s(\th), \qquad
\F^\s(\aa)=\sum_{k-1}^\io \e^k \F^{\s,(k)}(\aa)\Eq(3.9)$$
we see that the gradient with respect to $\aa$ of $\F^{\s,(k)}(\aa)$ is
precisely $\AA^{a,(k)}(\aa)$. And the splitting $\V
Q(\aa)$ is the gradient of $\F(\aa)=\F^{+}(\aa)-\F^{-}(\aa)$.

One can get directly a graphical representation of $\F^{(k)}$ as:

$$\F^{(k)}(\aa)=\sum_{\th:\,k}\ig_{+\io}^{-\io} dt\,
{\rm Wal}_{\s(t)}(\th)\Eq(3.10)$$
where $\s(t)=\sign(t)$ and ${\rm Wal}_{\s(t)}(\th)$ is just the {\it
integrand} in the $\II_{\s,0}$ integral with respect to the first node
variable $\t=\t_{v_0}$ appearing in the evaluation of
$\Val_{\s}(\th)$. This concludes the construction of Eliasson's
potential.
%\ifnum\mgnf=0\pagina\fi

\*
\0{\bf\S4. Properties of the potential.}
\numsec=4\numfor=1\*

Many properties of the gradient $\V Q(\aa)=\dpr_\aa \F(\aa)$ have been
studied in [G3], [GGM1], [GGM2], [GGM3]: they are immediately
translated into properties of the potential $\F$, either by
integration or by following the proofs of the corresponding statements
for $\V Q(\aa)$.  We just summarize them:

(1) If $\oo$ is fixed then, generically,
the first order dominates:

$$\F(\aa)=\e\ig_{+\io}^{-\io} d\t\, f(\aa+\oo t,\f^0(t)) + O(\e^2)\Eq(4.1)$$
this is the well known Melnikov's result. We shall say that there is
``dominance of Melnikov's term'' for some quantity every time that the
lowest order perturbative term gives the dominant asymptotic behavior
for it in a given limiting situation. Hence in \equ(4.1) domination
refers to $\e\to0$.

In the ``one dimensional'' case not explicitly treated above, but much
easier, of a periodic forcing in which there is only one angle $\aa$
and one action $\AA$, Melnikov's domination remains true {\it even} if
the parameter $g$ becomes small provided the $\e$ is chosen of the
form $\m g^q$ for some $q>0$ (proportional to the degree $N_0$ of $f$
as a trigonometric polynomial in $\f$) and $|\m|$ small enough.

This is somewhat nontrivial: in [G3] there is a proof based on the
above formalism; other proofs are available as the problem is
classical. The nontriviality is due to the necessity of showing the
existence of suitable cancellations that eliminate values of graphs
contributing to $\F(\aa)$ higher order ``corrections'' (corresponding
to special graphs) which are individually present and, in fact, larger
than the first order contributions.

(2) The next case to study is the same case of $g$ small but with the
Hamiltonian \equ(2.1) (\ie quasiperiodically, rather than
periodically, and rapidly forced): let $g^2=\h$ and $\h<1$ be a
parameter that we want to consider near $0$. In this case, too,
convergence requires that $\e=\m\h^q$ for some $q>0$ (proportional to
the degree $N_0$ of $f$ as a trigonometric polynomial in $\f$) and
$|\m|$ small enough.

The problem is discussed already in [G3] and, following it, we consider
the graphs $\th$ that contribute to $\F^{(k)}$ and at each node we
decompose $f_{\d}$ into Fourier harmonics $f_{\d}(\aa,\f)=\sum_{\nn}
f_\nn(\f) e^{i\nn\cdot\aa}$. This leads to considering {\it new graphs}
$\th$ in which at each node $v$ a label $\nn_v$ is added signifying
that in the evaluation of the graph value the functions
$f_{\d_v}(\aa,\f)$ are replaced by $f_{\d_v,\nn}(\f)
e^{i\aa\cdot\nn}$. Of course at the end we shall have to sum over all
the ``momentum labels'' $\nn_v\in Z^2$. We call $F^{(k)}_{\th}$ the
contribution to $\F^{(k)}$ from one such more decorated graph. Then
from \S8 of [G3] one sees that:

$$\eqalign{|\F^{(k)}_\th|\,\le&\, D\, B^{k-1} \,k!^{\,p}\,
\big(\sum_{\nn'}
e^{-|\nn'\cdot\oo|\fra\p{2g}}\big)\,\prod_{v}|f_{\nn_v}|\cr
|\F^{(k)}_\nn|\,\le& (b \h^{-q})^{k}\cr}\Eq(4.2)$$
where $g=\hdp$ the sum over $\nn'$ runs over the {\it nonzero} values
of the sums of subsets of $\nn_1,\ldots,\nn_k$. The constants $B,D$
are bounded by an inverse power of $\h$ and $p>0$ is constant
(depending on the degree of $f$ in $\f$); the constants $b,q$ can be
bounded only in terms the maximum of $|f|$ in a strip $|\Im \a_j|,|\Im
\f|<\x$ on which the maximum is finite. The first property follows from the
analysis in \S8 of [G3]; the second is symply the statement that the
stable and unstable manifolds are analytic function of $\e$ with
radius of convergence proportional to $\h^q$ for some $q$ (essentially
a result of Graff, see \S5 of [CG]).

(3) A consequence of \equ(4.2) is that $\F(\aa)$ can be represented
as, see [GGM3] for details on the corresponding statement for the
gradient of $\F$:

$$\eqalignno{
\F(\aa)=&\e\sum_\nn e^{i\aa\cdot\nn}(M_\nn+\e D_\nn(\aa,\e))\cr
M_\nn=& \sum_n f_{\nn,n}\ig_{+\io}^{-\io} (1-\fra{\dot\f^0(t)}{\cosh
gt})\,\cos n\f^0(t)\,dt&\eq(4.3)\cr
|\dpr_\aa^h D_\nn(\aa)|<& \h^{-p_h}C_h |\nn|^h,\qquad p_h,q_h>0\cr}
$$
which allows us to say, very easily, that the Eliasson's potential is
``in some sense'' dominated by Melnikov's value at least in the
special cases in which $f_{\nn,n}$ are positive and ``{\it as large as
possible}'', \ie $ f_{\nn,n}= c\,e^{-\k|\nn|}\d_{n,1}$ for $c>0$, and
$\oo$ has good Diophantine properties, \eg if $\o_1/\o_2$ is the
golden mean (here $|\nn|\defi|\n_1|+|\n_2|$).

In the latter instance one verifies that, for all $\aa$,
$|D_\nn(\aa,\e)|<\h^{-q'} M_\nn$ for all $|\nn|<\h^{-1}$ which togheter
with the analyticity in $\e$ of $\F(\aa)$ allows disregarding the
contributions to $\F(\aa)$ from the $\nn$'s exceeding $\h^{-1}$. The
properties of the golden mean allow us immediately to see that in the
sum only one pair $\pm\nn$ dominates at $\aa=\V0$: it is the pair
$\nn_0=(f_k,-f_{k+1})$ if $f_j$ is the Fibonacci's sequence and $k$ such
that $\k |\nn_0|+\fra\p{2\hdp}|\oo\cdot \nn_0|$ is minimum; {\it apart}
from exceptional intervals of values of $\h$ in correspondence of which
there may be two pairs (or more) (see \S2,6 of [DGJS]). The domination
persists for all the $\aa$'s such that $|\sin\nn_0\cdot\aa|>b$ where
$b>0$ is any prefixed constant (the smaller $b$ the smaller has $\e$ to
be to insure dominance).

Also the gradient of $\F(\aa)$, and in fact any derivative of $\F$ is
dominated by the Melnikov's term, by the same type of argument. But
this is somewhat trivial: the real question is, in view of the remarks
in \S2 about the possible applications to heteroclinic strings and to
Arnold's diffusion, whether the homoclinic splitting is dominated by
Melnikov's integral. This seems to be, in the generality  considered
here, {\it still an open problem}. The reason is very simple; from
\equ(4.3) one easily deduces that:

$$\D=-\e^2\sum_{\nn,\nn'}
e^{-(|\oo\cdot\nn|+|\oo\cdot\nn'|)\fra\p{2}\hdm}
\big((\nn\wedge\nn')^2 M_\nn M_{\nn'}+ \e
d_{\nn,\nn'}\big)\Eq(4.4)$$
{\it hence one realizes that the term that should be leading,
$\nn=\pm\nn'=\pm\nn_0$}, is missing in the lowest order
part. Therefore the main contribution comes, {\it or may come}, from
the remainder $d_{\nn,\nn'}$ on which we have little information
besides the above bounds (which would be plenty if the Melnikov's main
term did not vanish). Curiously the above exceptional cases, \ie when
the value of $\h$ is taken along a sequence $\h_j\to0$ such that for
each $j$ there are {\it two} minimizing vectors $\nn_0$ and $\nn'_0$,
can be, instead, easily solved because $(\nn_0\wedge\nn'_0)^2\ge1$ as
no two Fibonacci's vectors can be parallel.

In the literature there are various claims about ``proofs'' of
dominance of Melnikov's contribution to the splitting: they however
seem to be always proofs of the ``easy part'' namely of the dominance
of the Melnikov's term in {\it some components of} the splitting vector
$\V Q(\aa)$ (implied by the above analysis).

The only known case of generic dominance of the Melnikov term {\it for
the splitting} is the one discussed in [GGM1], see (5) below. And its
analysis is already far more subtle than the above.

(4) In general the estimates \equ(4.2), called ``quasi flat'' in [G3]
are {\it optimal} (see [GGM4]): hence one {\it cannot hope} to have
bounds on the Fourier transform of $\F$ of the form, for some $r>0$:
$|\F_\nn|< const\, \h^{-r} e^{-\fra\p{2g}|\oo\cdot\nn|}
e^{-\k|\nn|}$. Such estimates are called ``exponentially small'' and,
occasionally, have been claimed to be possible.

(5) The above results are {\it very easy} compared to the ones that
can be obtained by taking $g^2$ {\it fixed} and $\oo=(\hdp \o,\hdm
\o')$, discussed in [GGM1] and called the {\it three time scales
problem}, because the system has three time scales of orders
respectively $\hdm,1,\hdp$. In this case we consider the values of
$\h$ for which $\oo$ verifies a Diophantine property of the form
$|\oo\cdot\nn|>\h^\g|\nn|^{-t}$ with some $\g,\t>0$ and we take $\e$
equal to a suitably large power of $\h$ so that the small divisors
problems can be overcome and the invariant tori do exists.

The quasi flat estimates hold (for small $\h$) but they {\it do not
imply that the matrix $\dpr_{\aa,\aa}\F(\aa)|_{\aa=\V0}$ has three
matrix elements of size exponentially small as $\h\to0$}. In fact all
the four matrix elements are of the order of a power of $\h$: this is
so {\it in spite of the fact that the Melnikov term $M(\aa)$
generates a contribution to the $2\times2$ splitting matrix with three
exponentially small entries}.

In other words neither $\F$ nor the Hessian matrix
$\dpr_{\aa,\aa}\F|_{\aa=\V0}$ are dominated by the the Melnikov's
``first order'' contribution. {\it Nevertheless the Melnikov's
contribution to the Hessian gives the leading term in the limit
$\h\to0$}! (generically in the perturbation). 

Of course if the above mentioned exponential estimates could be
correct this would follow immediately from them: but they are not
valid (as they would imply the wrong statement that the splitting
matrix has $3$ exponentially small entries) and the result holds {\it
only because remarkable cancellations take place}. Hence, contrary to
what is sometimes stated, the above case requires a delicate analysis,
compared to the one in [G3] which solves easily the problems
(1)$\div$(4) above at least as far as the domination of the first
order in the derivatives of the Eliasson's function (hence the
splitting vector) is concerned.  In particular this means that $\F$
{\it is not a good measurement} of the splitting.

It is in the theory of this ``three time scales problem'' that the
analogy with field theory and renormalization theory turns out to be
particularly useful and the methods characteristic of such theories
apply very well and turn into a rather simple matter the check of the
infinitely many identities that are necessary in order that all terms
in the Hessian that dominate the Melnikov contribution cancel each
other leaving out just the Melnikov contribution as the leading one as
$\h\to0$.

(6) The just described graphical technique seems not only very well
suited for the questions analyzed or mentioned above but it seems
quite promising also with respect to the solution of one of the main
standing problems, namely: what is the asymptotic behavior of the
splitting as $g\to0$ and $\oo$ fixed?  The case mentioned in (3) above
requires that all the Fourier components of the perturbation do not
vanish: a finer analysis shows that this can be somewhat weakened but
{\it not} to the extent of allowing polynomial perturbations. Hence
such cases seem to have a rather limited interest: they appear in fact
too special. But even so the only thing we know is the Melnikov's
dominance in Eliasson's potential and in its derivatives.

On the other hand the theory discussed in [G3], [GGM3], and in \S3
suggests the following question. First of all let us {\it define} an
extension of Melnikov's function to {\it higher order}. We simply
consider the function $\F^0(\aa)$ which is obtained from the
diagrammatic representation \equ(3.9) {\it but replacing the operators
$\OO$ associated with the nodes of $\t$ by the operator}:

$$\OO_0(F)(t)=\fra12\sum_{\ch=\pm}\ig_{\ch\io}^t
\big(w_{01}(t)w_{00}(\t)-w_{00}(t) w_{01}(\t)\big)\,
F(\t)\,d\t\Eq(4.5)$$
then, supposing $\oo_0$ with golden rotation number (or any number
with very good Diophantine nature):
\*
\0{\it Conjecture:  In model \equ(2.1) and assuming that $g=\hdp$, 
will the Hessian of $\F^0(\aa)$ give the leading asymptotics as
$\h\to0$ of the splitting determinant at $\aa=\V0$ ``generically'' in
$f$ ?}
\*

Here generic means both genericity in the space of trigonometric
polynomial pertubations of fixed degree (arbitrary) and in the space
of the analytic perturbations, possibly with the constraint that the
perturbation is of positive or negative type. However we require that
the perturbation be polynomial in the $\f$ variable, see (9) below. As
far as I know there is no proof even of the convergence of the series
defining $\F^0$ (which is well defined only as a formal series and
which may have to be regarded as an asymptotic series, see [G3], [GGM1]).

The conjecture can be extended to the case of three time scales considered
in (5): in that case it is affirmatively answered in [GGM], where,
however, one also sees that the Eliasson's potential and its derivatives
is {\it not} dominated by the first order. It is only the splitting
determinant that is dominated by the first order: {\it not surprisingly
as this is the only quantity among the ones discussed which has a direct
physical meaning}.

The conjecture could be strengthened by adding, for instance, that
$\F^0$ can be replaced by the function $\tilde F^0$ obtained from $\F^0$
by developing in powers of $\e$ its Fourier coefficients $\F^0_\nn$ and
retaining only the lowest non vanishing order $\tilde\F^0_\nn$ of each
Fourier coefficents to form the Fourier transform of $\tilde\F^0$. In
this stronger form it becomes, in the assumptions of (2) above (fast
forcing and ``maximal size of the Fourier coefficients of the
perturbation), simply the statement that the splitting determinant can
be computed by the first order Melnikov's integral: an open problem (as
mentioned above). However in this form the conjecture is not really
stronger than above because using $\F^0$ instead of $\F$ amounts to
saying that the $\e d{\nn,\nn'}$ in \equ(4.4) has the form $\e
(\nn\wedge\nn')^2 d'_{\nn,\nn'}$.

Clearly in order that the answer to the question be affirmative one
has to show the existence of suitable cancellations: I have checked
that {\it they are indeed present at the order beyond the lowest}
(since the lowest order for the homoclinic determinant is the second,
this means that the answer is affirmative to third order). The check
requires using the results in [GGM1], which may already imply a
positive answer to all orders.

Denoting $\OO$ the operator $\OO(F)(t)\defi\OO_{\s(t)}(F)(t)$ (with
$\s(t)=\sign(t)$) one remarks that in all the expressions involved in
the graphs evaluations one always really uses $\OO$; then it is useful
to note the (algebraic) relation between the operator $\OO$ and
$\OO_0$:

$$\eqalignno{
\OO F(t)=&\OO_0 F(t)+ |w_{01}(t)|\, G_0(F)+ w_{00}(t)\, G(F)\cr
\OO_0 F(t)=&\fra12\sum_{\ch=\pm}\ig_{\ch\io}^t
\big(w_{01}(t)w_{00}(\t)-w_{00}(t) w_{01}(\t)\big)\, F(\t)\,d\t&\eq(4.6)\cr
G_0(F)=& \fra12 \ig_{+\io}^{-\io} d\t\, w_{00}(\t)\, F(\t)\,d\t\,,
\qquad G(F)=
\fra12 \ig_{+\io}^{-\io} d\t\, |w_{01}(\t)|\, F(\t)\,d\t\cr}$$
and, as it is clear from [GGM1], the $G,G_0$ factors play the role of
``counterterms'' in the field theory interpretation of the diagrammatic
expansion of $\F$. Hence the above question suggests that the leading
behavior of the splitting determinant is due to graphs without
counterterm contributions.\annota5{Called in field theory ``most
divergent'' graphs: rather improper an expression because in any
reasonable field theory there should be no divergences at all; as it is
the case in the theories that have been actually shown to exist on a
mathematical basis} Here the ``counterterms'' contain non analytic
functions and they are responsible for the impossibility of
exponentially small estimates. A positive answer to the above question
would state that they only give rise to subleading contributions to the
splitting.

An explicit expression for the value contributing to $\F^0$ can be
found in [GGM1]: see (6.2), for the isochronous case \equ(2.1), and
see the paragraph preceding (7.4) for the anisochronous case.

(7) The above theory can be immediately extended to anisochronous
cases: one just has to consider a few new types of graphs, [G3], that
contribute to the splitting vector $\V Q(\aa)$ and to the splitting
potential $\F(\aa)$.

(8) Most of the considerations above do not really use that the
dimension of the quasi periodic motion is $2$: if it is suppoo=sed
larger it is however difficult to see what will be the leading
behavior of the splitting. One reason is that even the analysis of the
Melnikov term is itself a quite difficult task: Diophantine
approximation theory is in a very rudimentary stage if the dimension
of the quasi periodic motion is $\ge3$.

A glimpse of the difficulties that one should expect to meet is given
by the three time scales problem (5) above.  In this problem we can
think that the slow frequency of order $\hdp$ is in fact obtained
because the perturbation by a {\it three dimensional} quasi periodic
motion with three {\it fast } frequences $\o_1,\o_2,\o_3$ of order
$\hdm$ contains an almost resonant harmonic $\nn$ such that
$\n_1\o_1+\n_2\o_2=O(\hdp)$. One would then naively think that in this
case the homoclinic splitting can become ``large'' because we can have
$\oo\cdot\nn$ small of order $\hdp$ with not too large $\n_1,\n_2$. But
{\it this is illusory} precisely because from the results of the case (5)
one sees that, although we could expect a large splitting vector and
matrix, its Hessian at the homoclinic point will be exponentially
small as $\h\to0$. Therefore in the three dimensional case we should
expect that the resonances {\it do not enhance the splitting}: they
can make large the spliting matrix but not its determinant! This
remark also explains why the problem (5) above is so unexpectedly
difficult to analyze (see [GGM1]).

(9) Finally there seems to be no reason whatsoever for having a small
homoclinic splitting when the perturbation is not a polynomial (but
``just'' analytic) in the $\f$ variable, not even when the rotation
vector $\oo$ is very fast.
\*

\0{\it Acknowledgments: I am honored to have been asked by Luis
Michel to contribute to this volume. I am also grateful to the
Directors and Members of IHES who made possible the development of
many of my scientific works through the frequent invitations to visit
IHES in the last 33 years. For the technical part of this paper I am
greatly indebted to G.Gentile, V. Mastropietro, G. Benfatto,
G.Benettin, A. Carati: their suggestions and help have been
essential.}
\*
\vskip1truecm
\0{\bf References.}
\*
\vskip1truecm

\0[BCG] Benettin, G., Carati, A., Gallavotti, G.: {\it A rigorous
implementation of the Jeans--Landau--Teller approximation for
adiabatic invariants}, Nonlinearity {\bf 10}, 479--507, 1997.
\*

\0[CG]  Chierchia, L., Gallavotti, G.:
{\it Drift and diffusion in phase space},
Annales de l'In\-sti\-tut Henri Poincar\'e B {\bf 60}, 1--144, 1994.
\*

\0[DGJS] Delshams, S., Gelfreich, V.G., Jorba, A., Seara, T.M.:
{\it Exponentially small splitting of separatrices under fast
quasiperiodic forcing}, Communications in Mathematical Physics
{\bf189}, 35--72, 1997.
\*

\0[E] Eliasson, L.H.: {\it Absolutely convergent series expansions 
for quasi-periodic motions}, Ma\-the\-ma\-ti\-cal Physics Electronic
Journal, {\bf 2}, 1996.
\*

\0[Ge] Gelfreich, V.G.: {\it A proof of exponentially small
transversality of the sepratrices for the standard map}, in
mp$\_$arc@math. utexas. edu, \#98-270: this recent paper, besides
clarifyng various aspects of previous papers, provides an accurate
exposition of the main ideas (and appropriate references) of the other
papers by the russian school.
\*

\0[GGM0] G. Gallavotti, G. Gentile, V. Mastropietro: {\it Field theory
   and KAM tori}, p. 1--9, Mathematical Physics Electronic Journal,
   MPEJ, {\bf 1}, 1995 (http:// mpej.unige.ch), .
\*

\0[GGM1] G. Gallavotti, G. Gentile, V. Mastropietro: {\it Pendulum:
  separatrix splitting}, in mp$\_$arc@math.utexas.edu, \# 97-472.  To
  appear with a different title: {\it Separatrix splitting for systems
  with three time scales}.
\*

\0[GGM2] G. Gallavotti, G. Gentile, V. Mastropietro: {\it
  Hamilton-Jacobi equation, heteroclinic chains and Arnol'd diffusion
  in three time scales systems}, mp$\_$arc@math. utexas. edu \#98-4;
  chao-dyn@xyz. lanl. gov \#9801004.
\*

\0[GGM3] G. Gallavotti, G. Gentile, V. Mastropietro: {\it Melnikov's
approximation dominance. Some examples}, mp$\_$arc@math. utexas. edu
\#98-???; chao-dyn@xyz. lanl. gov \#9804043.
\*

\0[GGM4] G. Gallavotti, G. Gentile, V. Mastropietro: {\it Homoclinic 
splitting, II. A possible counterexample to a claim by Rudnev and
Wiggins on Physica D}, chao-dyn@xyz. lanl. gov \#9804017.
\*

\0[HM] Holmes, P., Marsden, J.: {\it }. See also:
Holmes, P., Marsden, J., Scheurle,J: {\it Exponentially Small
Splittings of Separatrices with applications to KAM Theory and
Degenerate Bifurcations}, Contemporary Mathematics, {\bf81}, 213--244,
1988.
\*

\0[T] Thirring, W.: {\it Course in Mathematical Physics}, vol. 1,
p. 133, Springer, Wien, 1983.
\*
\*

\0{\it Author's preprints at {\tt http://ipparco.roma1.infn.it}}

\0{\sl Address: Fisica, Universit\'a di Roma ``La Sapienza'', P.le Moro
2, 00185 Roma, Italy}

\0{e-mail:\tt \ giovanni@ipparco.roma1.infn.it}

\ciao